\documentclass[epsfig,12pt] {article}
\usepackage{graphics}
\usepackage{epsfig}
\usepackage{amssymb}
\usepackage{amsmath}
\usepackage{bm}
\usepackage{epsfig}

\begin{document}

\baselineskip .7cm

\author{ { Vishesh Kaushik$^{\dagger}$,\ Navin Khaneja \thanks{To whom correspondence may be addressed. Email:navinkhaneja@gmail.com}} \thanks{System and Control Engineering, IIT Bombay, Powai - 400076, India.}}

\vskip 4em

\title{\bf Feedback Pulses}

\maketitle

\vskip 3cm

\begin{center} {\bf Abstract} \end{center}
We have a new paradigm to design NMR pulses. Pulses, we call feedback pulses. We want broadband inversion and excitation. We have many offsets, start evolving them all starting from the north pole. Monitor them on the Bloch sphere, see which offset is worst (most away from south pole). Change the rf-phase to the offset ($\pi/2$ ahead of offset transverse magnetization phase) and irradiate at that offset frequency and evolve for some time and monitor and repeat, looking for worst offset. When we are on resonance to a offset, we are doing well, inverting it and when we are off resonant, we don't hurt much (even if hurt little, we will come back to the offset in good time). By the process of monitoring, and setting phase we eventually push everything to the south pole and bingo, we have an inversion pulse. Feedback is done in simulation, but what results in end is a broadband inversion pulse. For broadband excitation, start with all offsets (symmetric around origin) on y axis. By feedback push them to the south pole. When we run the resulting sequence backward with phases, $\pi$ incremented, we will get an excitation pulse. For band-selective excitation pulse put offsets in pass band on the $y$ axis and in the stop band on the south pole. Use feedback to push everything to the south pole. Again, run backwards with $\pi$ incremented phases, to get band selective excitation. Suddenly, we have it all, simple and easy. The paper, introduces the feedback pulse algorithm, simulations and experiments.

\vskip 3em

\section{Introduction}
The excitation and inversion pulse is ubiquitous in Fourier Transform-NMR, being the starting point of all experiments. With increasing
field strengths, in high resolution NMR, sensitivity and resolution comes with the challenge of
uniformly exciting or inverting larger bandwidths. At a field of 1 GHz, the target bandwidth is $50$ kHz for excitation of entire 200 ppm $^{13}$C chemical shifts. 
The required $25$ kHz hard pulse exceeds the capabilities of most
$^{13}$C  probes and poses additional problems in phasing the spectra. 
In $^{19}$F NMR, chemical shifts can range over 600 ppm, which requires excitation of different regions of the spectra. Methods that can achieve
uniform excitation over the entire bandwidth in $^{19}$F NMR, are therefore most desirable.
Towards this end, several methods have been developed for 
broadband excitation/inversion, which have reduced the phase variation of the excited magnetization as a function of the resonance offset. 
These include composite pulses, adiabatic sequences, polycromatic sequences, phase alternating
pulse sequences, optimal control pulse design, and method of multiple frames, \cite{comp1}-\cite{ultrabroadband}.

Another application is frequency (band) selective excitation. Frequency-selective pulses have widespread use in magnetic resonance
and significant effort has been devoted towards their design \cite{barrett}-\cite{glaser}.
Several experiments in high-resolution NMR and magnetic resonance
imaging require radio-frequency pulses which excite NMR response over a prescribed frequency range with negligible effects elsewhere.
Such band-selective pulses are particularly valuable when the excitation is uniform over desired bandwidth and of constant phase.

In this paper, we introduce a new paradigm for NMR pulse design, the concept of feedback pulses. In classical control, feedback is the way to go. Whether it is regulating speed of a car, or opening a control valve by certain amount, or setting temperature in a room or changing missile course by $5^{\circ}$ or steering a rocket to moon, feedback is the default design method. The parameters of a system dynamics are not always exactly known and in order to steer the system to a target state, we don't apply a known control pulse from a library, instead we monitor the system and measure its current value and nudge it in direction of target and keep doing it till it is at target. We use the same idea to design broadband excitation and inversion pulses. We want broadband inversion and excitation. We have many offsets, start evolving them all starting from north pole. Monitor them on Bloch sphere, see which offset is worst (most away from south pole). Change the rf-phase to the offset and irradiate at that offset frequency and evolve for some time and monitor and repeat, looking for worst offset. When we are on resonance to a offset, we are doing well, inverting it and when we are off resonant, we don't hurt much. By process of monitoring, setting phase we eventually push everything to south pole and we have a inversion pulse. For broadband excitation, start with all offsets (symmetric around origin) on y axis. By feedback push them to the south pole. When we run the resulting sequence backward with phases, $\pi$ incremented, we will get an excitation pulse. For band-selective pulse put offsets in pass band on the $y$ axis and in the stop band on the south pole. Use feedback to push everything to the south pole. Again run backwards, with $\pi$ incremented phases, to get selective excitation. The paper introduces this feedback pulse algorithm, simulation and experiments.

\section{Theory and Experiments}
\subsection{Algorithm}
Let $A$ be amplitude of rf pulse. Let $(z_j, \phi_j)$ be the $z$ coordinate and phase of transverse magnetization for the $j^{th}$ offset (where offsets lie in range $[-B, B]$).
Start with all $z_j = 1$ (north pole). Find $j$ with target $z_j$ (most away from south pole), change rf pulse phase to $\theta_i = \phi_j + \frac{\pi}{2}$ and give a pulse of small flip angle $\alpha = \frac{1}{2}$ degree and repeat and continue till all $z_j$ are near south pole then stop. Then $\theta_i$ is the broadband inversion pulse.

Let $(z_j, \phi_j)$ be the $z$ coordinate and phase of transverse magnetization for the $j^{th}$ offset.
Start with all $z_j = 0, \phi_j = 90^{\circ} $ (y axis). Find $j$ with largest $z_j$ , change rf pulse phase to $\theta_i = \phi_j + \frac{\pi}{2}$ and give a pulse of small flip angle say $\alpha = \frac{1}{2}$ degree and repeat and continue till all $z_j$ are near south pole then stop. Then $\theta_i = \theta_{n-i} +180^{\circ}$ is the broadband x-excitation pulse, {\bf as the offset $\omega$ will backtrack the trajectory of $-\omega$ and go back from pole to equator.}

Let $(z_j, \phi_j)$ be the $z$ coordinate and phase of transverse magnetization for the $j^{th}$ offset (where offsets lie in range $[-B, B]$, with $[-C, C]$ as pass band and outside stop band).
Start with all $z_j = 0, \phi_j = 90^{\circ}$ (y axis) for pass band and  $z_j = -1$ (south pole) for stop band. Find $j$ with largest $z_j$ , change rf pulse phase to $\theta_i = \phi_j + \frac{\pi}{2}$ and give a pulse of small flip angle say $\frac{1}{2}$ degree and repeat and continue till all $z_j$ are near south pole then stop. Then $\theta_i = \theta_{n-i} + 180^{\circ}$ is the band selective x-pulse.

\subsection {Simulations}

\begin{figure}[htp!]
  \centering
  \includegraphics[scale = .75]{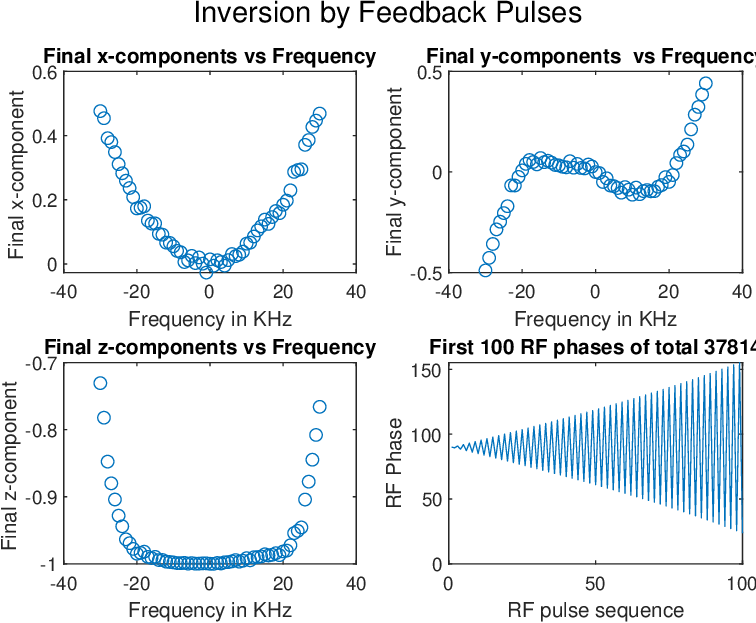}
  \caption{Fig. depicts inversion with feedback pulses over a bandwidth of $40$ kHz with rf amplitude $10$ kHz and time of pulse 3 ms.} \label{fig:inversion}
\end{figure}

\begin{figure}[htp!]
  \centering
  \includegraphics[scale = .75]{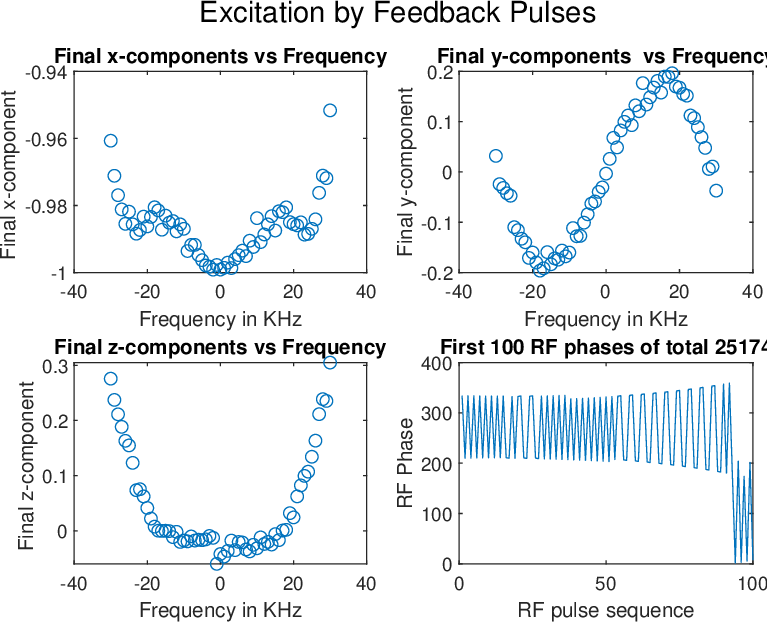}
  \caption{Fig. depicts excitation with feedback pulses over a bandwidth of $40$ kHz with rf amplitude $10$ kHz and time of pulse 2 ms.} \label{fig:excitation}
\end{figure}

In above, for broadband inversion and excitation, $A = 10$ kHz, $B = 20$ kHz, $\alpha = .57^{\circ}$. Fig. \ref{fig:inversion} and \ref{fig:excitation} shows $x$, $y$, $z$ coordinates as function of offset after application of broadband inversion and excitation. Also shown are the rf-phases. The two sequences take 3 ms and 2 ms respectively and have 18907 and 12587 number of  phases respectively. We used 40 offsets in the range.

\begin{figure}[htp!]
  \centering
  \includegraphics[scale = .75]{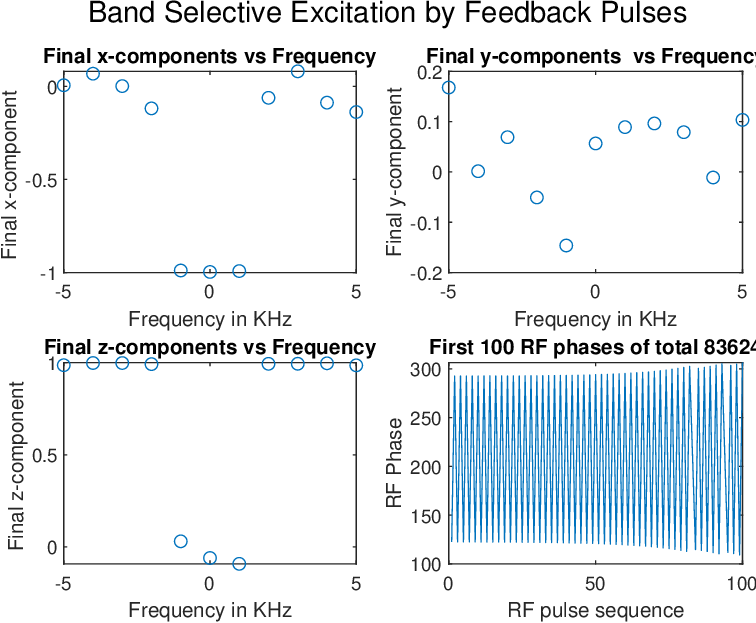}
  \caption{Fig. depicts selective excitation with feedback pulses over a bandwidth of $4$ kHz with rf amplitude $5$ kHz and time of pulse 6.7ms .} \label{fig:band}
\end{figure}

In above, for band selective excitation, $A = 5$ kHz, $B = 5$ kHz, $C = 2$ kHz, $\alpha = .29^{\circ}$. Fig. \ref{fig:band} shows $x$, $y$, $z$ coordinates as function of offset after application of selective excitation. Also shown are the rf-phases. The sequence takes 6.7 ms and have 41812 number of  phases.

\begin{figure}[htp!]
  \centering
  \includegraphics[scale = .75]{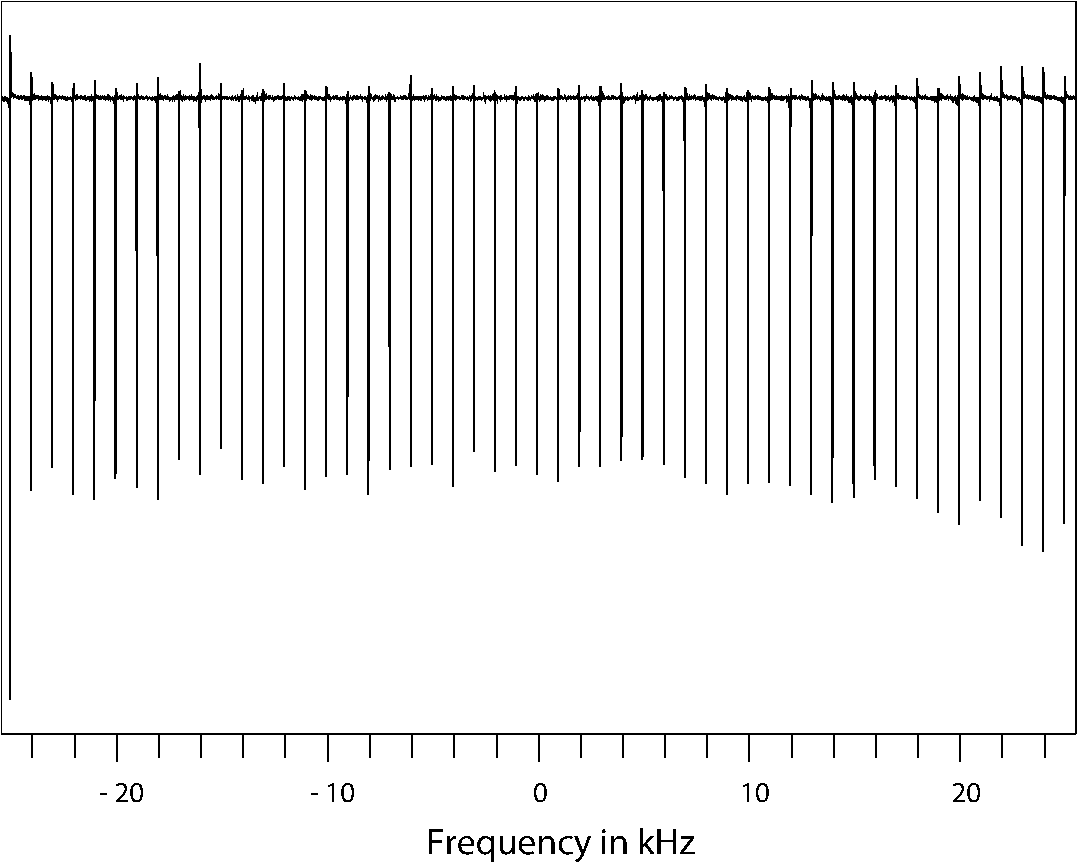}
  \caption{Fig. depicts 40 kHz wide inversion profile (HDO signal) of feedback inversion pulse on a sample of $99.5 \%$ D$_2$0 and$.5 \%$ H$_2$0. Rf power is $10$ kHz and pulse is 3 ms long.} \label{fig:einversion}
\end{figure}

\begin{figure}[htp!]
  \centering
  \includegraphics[scale = .75]{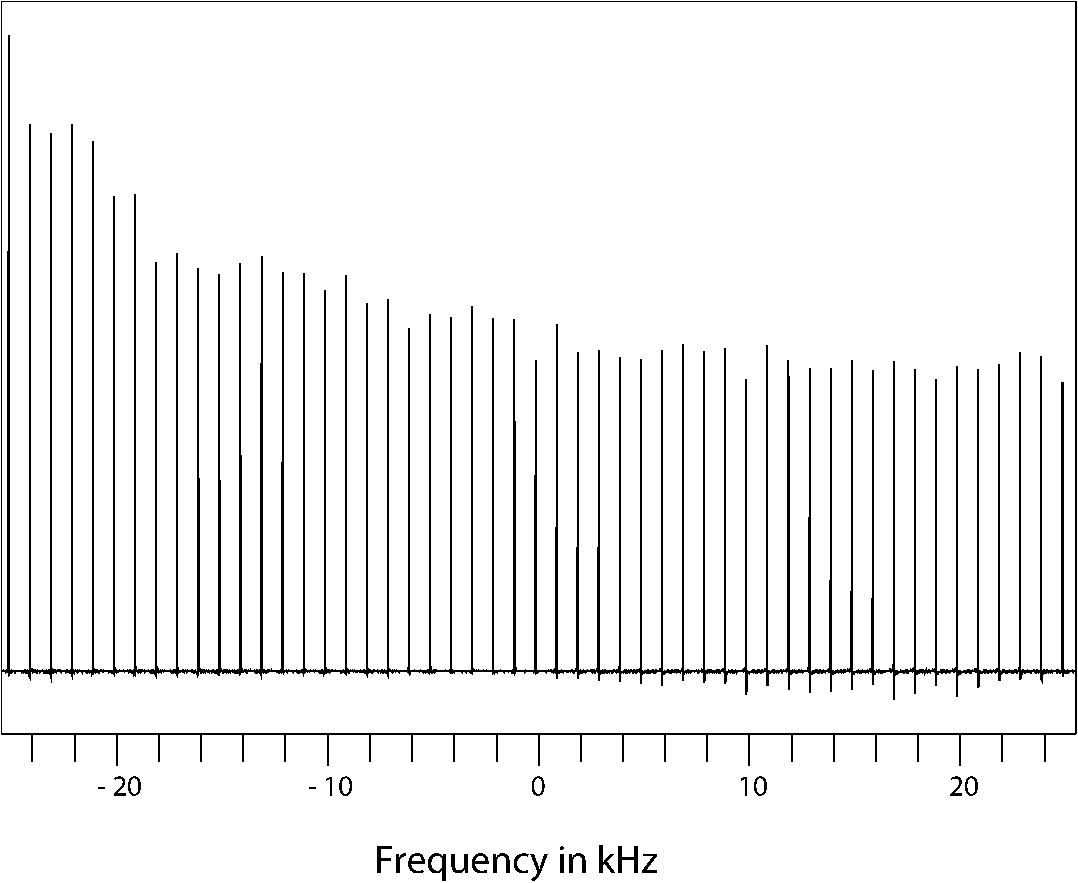}
  \caption{Fig. depicts 40 kHz wide excitation profile of feedback inversion pulse on  a sample of $99.5 \%$ D$_2$0 and$.5 \%$ H$_2$0. Rf power is $10$ kHz and pulse is 2 ms long.} \label{fig:eexcitation}
\end{figure}

\begin{figure}[htp!]
  \centering
  \includegraphics[scale = .75]{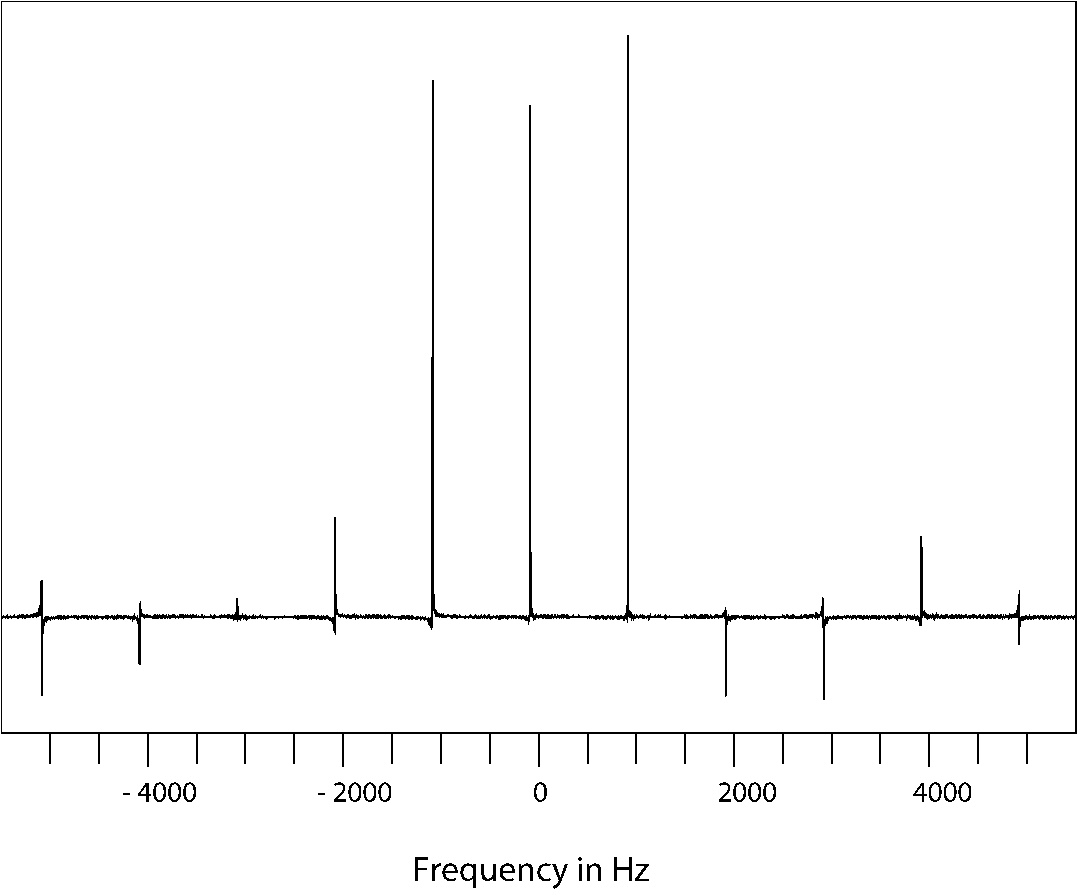}
  \caption{Fig. depicts 4 kHz wide selective excitation profile of selective pulse on  a sample of $99.5 \%$ D$_2$0 and$.5 \%$ H$_2$0. Rf power is $5$ kHz and pulse is 6.7 ms long.} \label{fig:eband}
\end{figure}

Based on simulations, experiments for broadband inversion, excitation and band selective excitation were performed on a $99.5 \%$ D$_2$0 and$.5 \%$ H$_2$0 solution. Water resonance at 4.7 ppm is excited as different offsets by varying transmitter frequency.
Fig. \ref{fig:einversion}, \ref{fig:eexcitation} and \ref{fig:eband}, shows experimental inversion, excitation and band selective profiles respectively.

\section{Conclusion}
In this paper, we introduced a new paradigm for design of NMR pulses, the pulses we call feedback pulses. 
By monitoring, the offsets and setting the phase of the rf-pulse to an offset, we can talk to it and drive it south pole and
repeat the process to engineer broadband inversion, excitation and selective excitation pulses. We provided algorithms, simulations and experiments to validate this methodology.

In classical control feedback is a popular method. Whether it is regulating speed of car, or opening a control valve, or setting temperature in a room or guiding a missile or steering a rocket to moon, feedback is the default design method. The parameters of a system dynamics are not always exactly known and in order to steer the system to a target state we don't apply a known control pulse from a library, instead we monitor the system and measure its current value and nudge it in direction of target and keep doing it till it is at target.

It is expected the feedback method proposed in this paper will find widespread use in other applications in NMR like frequency selective excitation with arbitrary profile with multiple bands, design of broadband propagators (like refocusing pulses) and rf-inhomogeneity compensation. In our implementation, we monitor the worst offset, in other and future implementations we may choose different polling strategy, by going through the offsets in a linear order (as in a chirp) and setting the phase to a offset as we go and poll it and repeating the process, till all offsets are on south pole.

\section{Acknowledgement}

We thank Prof. Ashutosh Kumar and Pramod Mali in HFNMR lab at IIT Bombay for their help with NMR data collection.

\end{document}